\documentclass[%
aip-cp,
 jmp,%
 amsmath,amssymb,
preprint,%
]{revtex4-1}

\usepackage{graphicx}

\usepackage{dcolumn}%
\usepackage{bm}%
\usepackage{neuralnetwork}
\usepackage{tikz}

\def\ltsima{$\; \buildrel < \over \sim \;$}
\def\gtsima{$\; \buildrel > \over \sim \;$}
\def\simlt{\lower.5ex\hbox{\ltsima}}
\def\simgt{\lower.5ex\hbox{\gtsima}}

\usepackage{color}

\begin{document}
\preprint{ }

 \title{Solving higher-order Lane-Emden-Fowler type equations using physics-informed neural networks: benchmark tests comparing
 soft and hard constraints}

 
\author{Hubert Baty$ $  \\ {Observatoire Astronomique, CNRS UMR 7550,
 Université de Strasbourg, 67000 Strasbourg, France \\ 
 hubert.baty@unistra.fr
  \quad \\ }}

\date{\today}%

\begin{abstract}

In this paper, numerical methods using Physics-Informed Neural Networks (PINNs) are presented with the aim to solve
higher-order ordinary differential equations (ODEs). Indeed, this deep-learning technique is successfully applied for solving different classes of singular 
ODEs, namely the well known second-order Lane-Emden equations, third order-order Emden-Fowler equations, and fourth-order Lane-Emden-Fowler
equations. Two variants of PINNs technique are
considered and compared. First, a minimization procedure is used to constrain the total loss function of the neural network, in which the
equation residual is considered with some weight to form a physics-based loss and added to the training data loss that contains the initial/boundary conditions. 
Second, a specific choice of trial solutions ensuring these conditions as hard constraints is done in order to satisfy the differential equation,
contrary to the first variant based on training data where the constraints appear as soft ones. Advantages and drawbacks of PINNs variants are highlighted. 
   
   \textbf{Kewords:} Deep learning; Neural networks;  Physics-Informed neural networks
   \end{abstract}



\maketitle

\clearpage

\section{Introduction}

Deep learning techniques based on Neural Networks (NNs) can be used to solve differential equations in the following way. Unlike classical NNs which are based on a minimization
procedure of the error between solutions predicted by the NN and a dataset of exact known solution values (called training data),
Physics-Informed Neural Networks (PINNs) enhance NNs by defining some other set of data (called collocation points) at which the estimated solution
must additionally ensure the equation. The convergence is obtained via a loss function which expression is a measure of the error (e.g. the mean squared error).
This approach benefits from the possibility to evaluate exactly the differential operators at the collocation points by
using automatic differentiation, and is facilitated by use of Python open source software libraries like Tensorflow or Pytorch.
Therefore, the loss function is composed of two terms, a first one taking into account training data and a second one for the physics-based
information (i.e. the differential equation). The introduction of PINNs in such form is generally attributed to Raissi et al (2017, 2019) and is called vanilla-PINNs.
Many PINNs-variants can be actually found in the literature and applied to a variety of many different equations,
including ordinary and partial differential equations (see reviews by Cuomo et al. 2022, and Karniadakis et al. 2021).

In vanilla-PINNs, the minimum training data generally consists in the boundary (or initial) conditions that are necessary for the existence of solutions.
Consequently, the latter can be considered as constraints imposed in a soft way in the training process.
A tutorial with benchmark tests of the vanilla-PINNs methods applied to ordinary differential equations (ODEs) can be found in Baty and Baty (2023) and references therein.
However, an alternate PINNs-variant is particularly interesting. Indeed, it is possible to use specific choice of trial functions in order to satisfy
exactly the differential equation at boundaries, thus via hard constraints (Lagaris et al. 1998). In the following, this second variant is also considered, that is hereafter called hard-PINNs.

In a previous paper (Baty 2023), we have investigated the advantages and drawbacks of the vanilla-PINNs for solving
the different classes of classical second order differential Lane-Emden equations arising in astrophysics. The present work aims to generalize
our previous work to similar but higher order equations, namely third-order and fourth-order Lane-Emden-Fowler like equations, having eventually
multiple singularities. In particular, in this work we focus on the comparison between
the two variants.

 The paper is organized as follows. In Sect. 2, we summarize the basics of the PINNs with the two variants.
 Section 3 presents the results for a well-known second-order Lane-Emden equation investigated in Paper 1 in order
 to explain the differences between the two variants. The results concerning the third and fourth order Lane-Emden-Fowler like problems are reported in Sect. 4.
 and Sect. 5 respectively. Finally conclusions are drawn in Sect. 6.


  \section{The basics of PINNs}
  
  In this section, we summarize the basic concepts underlying the PINNs technique for the two variants previously cited in the introduction.
  
   \subsection{The basics of vanilla-PINNs}
   
   \begin{figure*}[!t]
\centering
\includegraphics[scale=0.33]{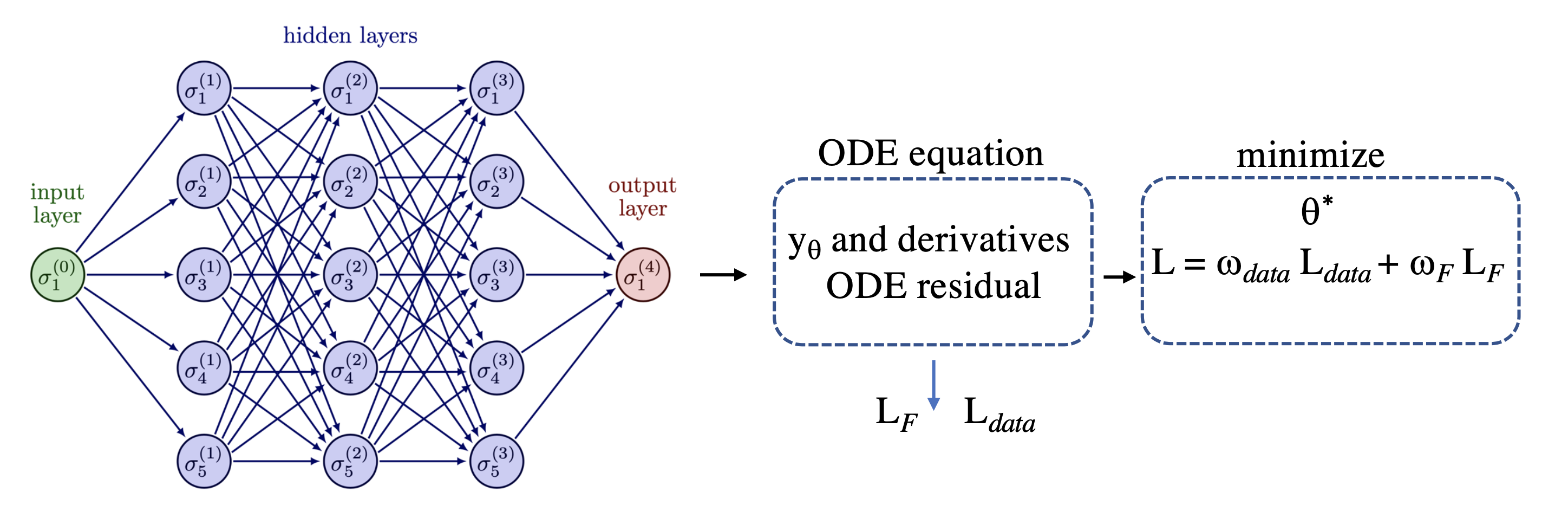}
  \caption{Schematic representation of network structure for a vanilla-PINN modelling an ODE. A NN architecture example is used to evaluate the residual of an ODE equation
  (via $y_\theta$ and associated higher order derivatives). Two partial loss functions are used to form a total loss function with associated weights (see text) that is finally minimized. 
  In this example, the input layer (single neuron) represents the $x$ variable and the output layer (single neuron) is the predicted solution $y_\theta (x)$, and $3$ hidden layers
  with $5$ neurons per layer are chosen.}
\label{fig1}
\end{figure*}   

 In the PINNs approach, we consider a dataset called collocation data
 aiming at evaluating the residual of the equation than can be written as
 \begin{equation}
   \mathcal{F} \left [x, y(x), y_x (x), y_{xx} (x), ... \right ] = 0,   \ \ \ \ \ \     x \in  \left[ 0,D \right] ,
\end{equation}
for a $k$ th-order ODE as considered in this work ($k$  values varying between $2$ and $6$). We also use the notation $y_x = \frac {dy} {dx}$,  $y_{xx}  = \frac {dy_x} {dx}$, ..., with $D$
defining the right boundary of the integration interval which may vary from case to case in this work.

A loss function associated to the physics (i.e. the equation residual) and called hereafter the physics-based loss can be defined as
\begin{equation}
    \mathcal{ L}_{ \mathcal{F}} (\theta) = \frac  {1} {N_c} \sum_{j = 1}^{N_c}  \left|   \mathcal{F} [ x_j, y_\theta (x_j), y_{x, \theta}(x_j),  y_{xx, \theta}(x_j), ... ]   \right| ^2 ,
\end{equation}
that must be evaluated at a set of $N_c$ data points located at $x_j$ (generally called collocation points, $j \in [1, N_c]$). 
 The exact solution $y(x)$ must be distinguished from $y_\theta (x)$ which is
 the corresponding approximated solution obtained via the neural network (i.e. predicted solution) for a set of model parameters defined by $\theta$ (see Baty and Baty 2023 for a tutorial).
Note that, 
the collocation data are arbitrary chosen and are not necessarily coinciding with the training data (see below).
As an important property characterizing PINNs, the different order derivatives of the expected solution with respect to the variable $x$ (i.e the NN input) needed in the previous
loss function are obtained via the automatic differentiation, avoiding truncation/discretization errors inevitable in traditional numerical methods.
For the vanilla-PINNs, in order to form a total loss function $ \mathcal{ L}  (\theta) $, the above physics-based loss function   $\mathcal{ L}_{ \mathcal{F}} (\theta)$
is added to a second loss function $ \mathcal{ L}_{data} (\theta)$
related to training data representing the solution knowledge at the boundaries. More precisely we have,
 \begin{equation}
           \mathcal{ L}  (\theta)    =   \omega_{data}  \mathcal{ L}_{data} (\theta)  +  \omega_{\mathcal{F}}  \mathcal{ L}_{ \mathcal{F}} (\theta),
\end{equation}
where weights (also called hyper-parameters) $(\omega_{data}, \omega_{\mathcal{F}})$ are introduced in order to ameliorate the eventual unbalance between
the two partial losses during the training process. 
These weights and the learning rate can be user-specified or automatically tuned. In the present work, for simplicity we fix the $\omega_{data} $ value to be constant and equal to unity, and the other weight parameters are determined with values varying from case to case.
More technical details about the PINNs methods can be found elsewhere (see Baty and Baty 2023 and references therein).
Obviously, the training data loss is defined  using a standard mean squared error formulation as,
  \begin{equation}
   \mathcal{ L}_{data} (\theta) = \frac  {1} {N_{data} } \sum_{i=1}^{N_{data} } \left|\ y_\theta (x_i ) - y_i^{data} \right|^2 .
\end{equation}
The latter expression assumes that a set of $N_{data}$ data is available for $y(x)$ taken at different $x_i$, i.e. input/output pairs ($x_i, y_i^{data}$)
are known that are generally called training data in the literature. In the present work, the training data are reduced to the sole knowledge of boundary
solution values, i.e. $y(0)=y_0$ and eventually to $y(D)=y_D$, thus $N_{data}  = 1$ or $2$.  In case of need of the additional knowledge of different derivatives at boundaries, the previous
data loss is modified as
 \begin{multline}
   \mathcal{ L}_{data} (\theta)  =  
    \frac  {1} {N_{data} } \sum_{i=1}^{N_{data} } \left|\ y_\theta (x_i ) - y_i^{data} \right|^2 
    +  \omega_d  \left[   \sum_{i=1}^{N_{data1} }      \left|\ y_{x, \theta}(x_i ) - y_{x,i}^{{data1}}     \right|^2    \right. \\
    \left.  + \sum_{i=1}^{N_{data2} }     \left|\ y_{xx, \theta}(x_i )  - y_{xx,i}^{{data2}}     \right|^2  + ...   \right],
 \end{multline}
where $N_{data1} $, $N_{data2} $, ..., stand for the number of first order, second order, ..., respectively known derivatives imposed at $x_i = 0$ or/and $ x_i = D$ which exact values are $ y_{x,i}^{{data1}}$,
$ y_{xx,i}^{{data2}}$, ..., respectively. An additional arbitrary weight factor $ \omega_d$ is also introduced. In fact, the above formula is the extension of the loss function definition proposed
previously in Baty (2023) for the second order Lane-Emden like equations. Of course, the first (and eventually the last) collocation point need to coincide with the training data point, as now 
the training data loss function include the soft constraints using the collocation points.

Finally, a gradient descent algorithm is used until convergence towards the minimum is obtained for a predefined accuracy (or a
given maximum iteration number) as
\begin{equation}
             \theta_{i+1} =  \theta_{i} - \eta  \nabla_{\theta}   \mathcal{ L}  (\theta_i) ,
\end{equation}
for the $i$-th iteration also called epoch in the literature,
leading to $  \theta^{*}  = \operatorname*{argmin}_\theta   \mathcal{ L}  (\theta)$, where $\eta$ is known as the learning rate parameter.
This is the so-called training procedure.
In this work, we choose the well known $Adam$ optimizer. The standard automatic differentiation technique is necessary to compute derivatives
(i.e. $\nabla_{\theta}$) with respect to the NN parameters (e.g. weights and biases).
The final goal of the method is to calibrate the trainable parameters $ \theta$ (weight matrices and bias vectors) of the network such that $y_\theta (x)$
approximates the target solution $y(x)$. Once trained, the NN directly gives an approximation that can be also written as
 \begin{equation}
y_\theta (x) =    \mathcal{NN} (x)   = ( \mathcal{N}_L \circ \mathcal{N}_{L-1} ...\  \mathcal{N}_0) (x) ,
\end{equation}
where the operator $\circ$ denotes the composition and $\theta =  \lbrace \boldsymbol{W}_l,  \boldsymbol{b}_l  \rbrace_{l =1,L}$ represents the trainable
parameters (weight matrices and bias vectors) of the network. The latter is composed of $L+1$ layers including $L-1$ hidden layers
of neurons, one input layer, and one output layer (as schematized in Fig. 1).
For each hidden layer we have,
\begin{equation}
 \mathcal{N}_l (x) =    \sigma ( \boldsymbol{W}_l  \mathcal{N}_{l-1} (x) +  \boldsymbol{b}_l ) ,
\end{equation}
where we denote the weight matrix and bias vector in the $l$-th layer by $\boldsymbol{W}_l  \in \mathbb{R}^{d_{l-1}  \times d_l}  $  
and $\boldsymbol{b}_l   \in \mathbb{R}^{d_{l} }$ ($d_l$ being the dimension of the input vector for the $l$-th layer). $\sigma(.)$ is a non linear
activation function, which is applied element-wisely. Such activation function allows the network to map nonlinear relationship that is
fundamental for automatic differentiation and therefore the calculation of the derivatives (see below).
In this work, me choose the most commonly used hyperbolic tangent $tanh$ function.

  \subsection{The basics of hard-PINNs}

Following the idea initially proposed by Lagaris et al. (1998), it is possible to use a trial function approach as
\begin{equation}
y_\theta (x) =   \mathcal{A} \left[ \mathcal{NN} (x) ,x  \right]  + \mathcal{B} (x) ,
\end{equation}
where $ \mathcal{NN} (x)$ is the NN approximation obtained by minimization of the physics-based loss function of the residual equation alone, and $\mathcal{B} (x)$ is
a well behaved smooth function that is chosen to satisfy the sole boundary conditions at $x = 0$, and $D$.
Consequently,  $\mathcal{B} (x)$ does not contain adjustable parameters. This is not the case of
the last function $\mathcal{A}$ that is constructed so as to not contribute to the above boundary conditions. The exact form of the above functions also depend on the
type of  boundary conditions (see below).

\section{Application to a second-order Lane-Emden equation}

We illustrate the application of the two PINNs variants on the well known second-order equation representing a polytropic
model of a gas sphere, arising in the theory of stellar structure in astrophysics (see Baty 2023 and references therein),
    \begin{equation}
     \frac {1}  {x^2}   \frac {d} {dx}   \left ( x^2  \frac {dy (x)} {dx}  \right ) + y^n = 0 ,
         \end{equation}
where $n$ is a positive polytropic index. The variable $x$ is a dimensionless radius, and the solution $y (x)$ is 
a normalized quantity related to the mass density. The index $n$ comes from the polytropic equation of state.
In this work, we consider only the value $n = 1$ for which the exact value exist that is
 \begin{equation}
     y (x) =  \frac {\sin(x)}  {x}  ,
    \end{equation}
in correspondance with the two boundary conditions at the origin $x = 0$, $y (0) = 1$ and
$\frac {dy (0)} {dx} = 0$. The corresponding exact first order derivative is,
 \begin{equation}
     y (x) =  \frac {\cos(x)}  {x}  - \frac {\sin(x)}  {x^2} .
    \end{equation}
Following Baty (2023), it is convenient to reformulate the differential equation using the equivalent form
   \begin{equation}
         x \frac {d^2y (x)} {dx^2}  +  2  \frac {dy (x)} {dx}  + x y^n = 0 ,
         \end{equation}
that is the residual equation expression effectively used and minimized in our PINN algorithm (see Eq. 1). In this way, the well known singularity
at the origin (that is problematic in some traditional scheme like Runge-Kutta integration) is circumvented.

The use of vanilla-PINNs on this case has been previously investigated in Baty (2023). For example, the integration over the spatial domain $D = 5$
using $N_c = 20$ uniformly distributed collocation points is illustrated in Fig. 2, where the predicted solution, the loss and MSE functions, and absolute errors obtained
 are plotted. The other parameters used are $4$ hidden layers with $20$ neurons per layer, with weight parameters $ \omega_{data}=1$, $\omega_d = 1 \times 10^{-2}$,
 $\omega_{\mathcal{F}} = 2 \times 10^{-2}$, and a learning rate $\eta = 1  \times 10^{-4}$. The training process is also stopped after $48000$ iteration steps (i.e. epochs).
     \begin{figure}[!t]
\centering
 \includegraphics[scale=0.34]{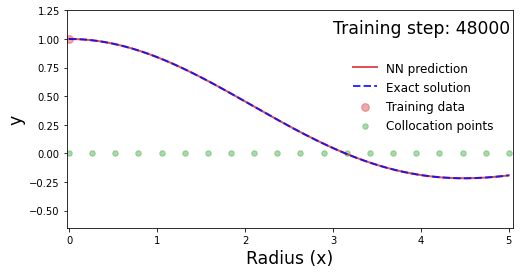}
  \includegraphics[scale=0.34]{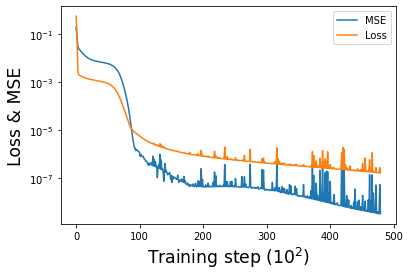}
   \includegraphics[scale=0.34]{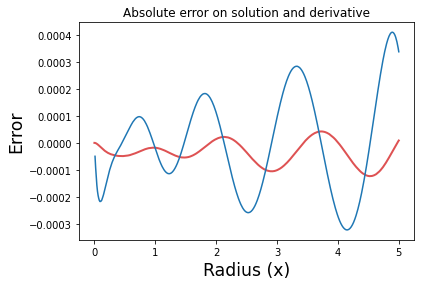}
  \caption{(Left panel) Predicted solution versus exact one using vanilla-PINNs obtained at the end of the training process for the second-order polytropic LE equation.
  The radius $x$ values for collocation points at which the physical loss function is evaluated are indicated with the green circles on $x$ axis. (Middle panel) The history
  of the total loss function and $MSE$ function during the training (see text). (Right panel) Absolute errors on solution (red) and first order derivative (blue) at the end of the training.
   }
\label{fig2}
\end{figure}   

    \begin{figure}[!t]
\centering
 \includegraphics[scale=0.34]{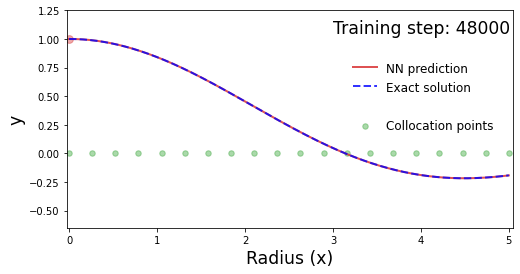}
  \includegraphics[scale=0.34]{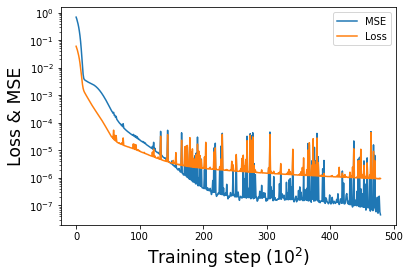}
   \includegraphics[scale=0.34]{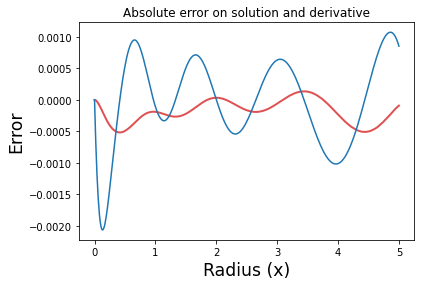}
  \caption{(Left panel) Predicted solution versus exact one using hard-PINNs obtained at the end of the training process for the second-order plytropic LE equation.
  The radius $x$ values for collocation points at which the physical loss function is evaluated are indicated with the green circles on $x$ axis. (Middle panel) The history
  of the loss function and $MSE$ function during the training (see text). (Right panel) Absolute errors on solution (red) and first order derivative (blue) at the end of the training.
   }
\label{fig3}
\end{figure}   

The MSE is evaluated using the standard
 expression, MSE  $=   \frac  {1} {N_{eval} } \sum_{i=1}^{N_{eval} } \left|\ y_\theta (x_i) - y_i^{eval} \right|^2 $, where the evaluation $y_\theta (x_i)$ is done
 on $N_{eval} = 500$ points uniformly distributed within the whole space interval, and $y_i^{eval}$ is the expected exact solution at $x = x_i$.
 This dataset introduced to test the accuracy of the method must not be confused with the collocation dataset at which the loss function is
 evaluated and used to make the progress of the training.
As one can see in Fig. 2, in this case we obtain a maximum absolute error of order $1 \times 10^{-4}$ on the solution $y$ and of order $4 \times 10^{-4} $ on its first derivative.
Note that the predicted first order derivative is directly deduced by the neural network (via automatic differentiation) at any $x$ value in the range of training interval
once the training process is finished. This latter property is a clear advantage of PINNs over a traditional numerical scheme.

We consider now the use of hard-PINNs. As explained above, training data are not needed but a trial function must be specified. We have followed the general form proposed
by Lagaris et al. (1998) for such initial value problem. Indeed, it is straightforward to check that the choice,
 \begin{equation}
     y_\theta (x) =  1 + x^2 \mathcal{NN} (x) ,
    \end{equation}
perfectly matches the required conditions detailed previously.
The results using a neural network with similar chosen parameters as for vanilla-PINNs (with the exception of the weights $ \omega_{data}$ and $\omega_d$
 that are not used) are plotted in Fig. 3.
Note also that only one loss function (the physics based one) is now used.
Despite the exact predicted solution and first derivative values at $x= 0$ (see right panel of Fig. 3) resulting from the hard constraints, the numerical errors
are not smaller and are even slightly worst compared to those obtained for vanilla-PINNs, at least for the combination of parameters used to produce these figures.
One must note that other combinations of parameters (number of layers, number of neurons, weights, number of collocation points, learning rate) can  lead to
a better accuracy for hard-PINNs compared to vanilla-PINNs (see also results for other problems below).
Generally speaking, the training process which final aim is calibrating the trainable parameters also depends on the initialization of these parameters. 
Indeed, such initialization being random (or more precisely quasi-random), it also influences the training and consequently the results (see Baty 2023 for
some convergence of the results with hyper-parameters and architecture of the neural network).

\section{Application to third-order Emden-Fowler equations}

These third-order Emden-Fowler equations are detailed in Verma and Kumar (2020) and can be categorized into two types. In this work, we
consider one example of the first type and one example of the second type.

 \subsection{Example of the first type}

We consider the following third-order Emden-Fowler equation
    \begin{equation}
     \frac {6}  {x^2}    \frac {dy} {dx}  +   \frac {6}  {x}  \frac {d^2y} {dx^2} +  \frac {d^3y} {dx^3 } = 6 (x^6 + 2 x^3 + 10)  e^{-3y}  ,
         \end{equation}
for $x$ in the range $[0 , 1]$, and subject to the conditions at the origin $x=0$ that are $y(0) = 0$, $y'(0) = 0$, and $y''(0) = 0$
(the notations $y'$ and $y''$ stand for first order and second order derivatives of $y$ respectively). Such first type equations are characterized
by double singularity, i.e. at $x = 0$ and $x^2 = 0$. Multiplying the different terms by $x^2$, the equation can be recast in the following form used for the residual form used below,
    \begin{equation}
     6 \frac {dy} {dx}  +   6x  \frac {d^2y} {dx^2} +  x^2 \frac {d^3y} {dx^3 } - 6x^2 (x^6 + 2 x^3 + 10)  e^{-3y} = 0 .
         \end{equation}
The analytic solution of the above equation is Ln$(1 + x^3)$.

We have used our two PINNs variants, where the trial function chosen for hard-PINNs is now
 \begin{equation}
     y_\theta (x) =  x^3 \mathcal{NN} (x) .
    \end{equation}
The results obtained using vanilla-PINNs and hard-PINNs are plotted in Fig. 4 and Fig. 5 respectively. 
We use $N_c = 11$ collocation points.
The neural network parameters used are $2$ hidden layers with $20$ neurons per layer, with weight parameters $ \omega_{data}=1$, $\omega_d = 1 \times 10^{-2}$ (for vanilla variant).
$100$ points are used for the $MSE$ and errors evaluations.
 $\omega_{\mathcal{F}} = 5 \times 10^{-2}$, and a learning rate $\eta = 5  \times 10^{-4}$ are chosen. The training process is also stopped after $92000$ iteration steps (i.e. epochs).

     \begin{figure}[!t]
\centering
 \includegraphics[scale=0.34]{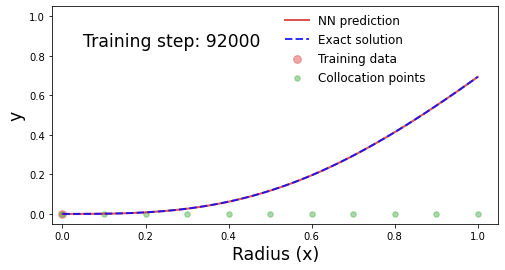}
  \includegraphics[scale=0.34]{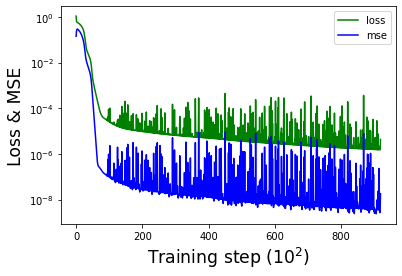}
   \includegraphics[scale=0.34]{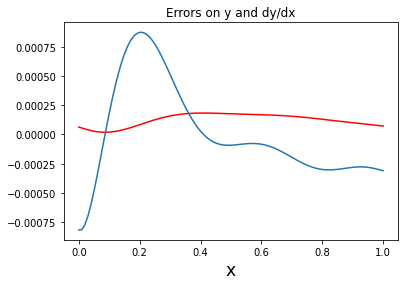}
  \caption{(Left panel) Predicted solution versus exact one using vanilla-PINNs obtained at the end of the training process for the third-order Emden-Fowler equation (first example).
  The radius $x$ values for collocation points at which the physical loss function is evaluated are indicated with the green circles on $x$ axis. (Middle panel) The history
  of the total loss function and $MSE$ function during the training (see text). (Right panel) Absolute errors on solution (red) and first order derivative (blue) at the end of the training.
   }
\label{fig4}
\end{figure}

     \begin{figure}[!t]
\centering
 \includegraphics[scale=0.34]{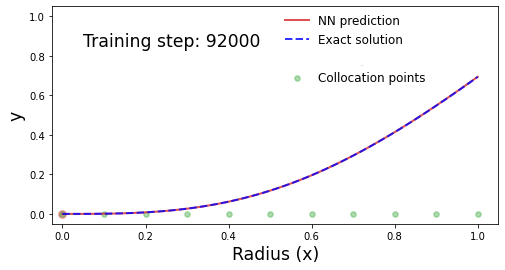}
  \includegraphics[scale=0.34]{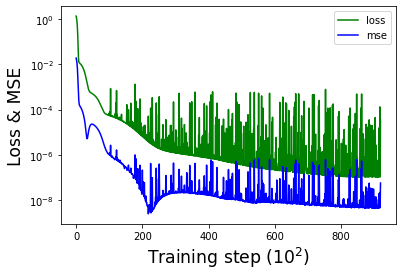}
   \includegraphics[scale=0.34]{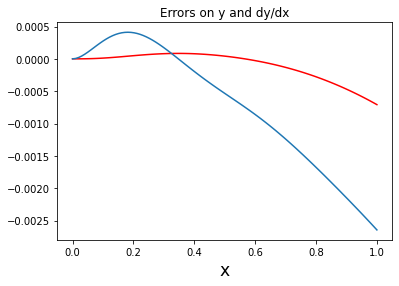}
  \caption{(Left panel) Predicted solution versus exact one using hard-PINNs obtained at the end of the training process for the third-order Emden-Fowler equation (first example).
  The radius $x$ values for collocation points at which the physical loss function is evaluated are indicated with the green circles on $x$ axis. (Middle panel) The history
  of the total loss function and $MSE$ function during the training (see text). (Right panel) Absolute errors on solution (red) and first order derivative (blue) at the end of the training.
   }
\label{fig5}
\end{figure}

\subsection{Example of the second type}

We consider now another example of the second type as follows (as also proposed by Verma and Kumar (2020) (in Problem 3.3),
and reformulated in a residual form as,
    \begin{equation}
       x \frac {d^3y} {dx^3 } -2  \frac {d^2y} {dx^2} - x y^3 - x( -x^9 e^{3x} + x^3 e^{x} + 7x^2 e^{x} + 6x e^{x}  -6 e^{x} ) = 0 ,
         \end{equation}
for $x$ in the range $[0 ,1]$, and subject to the conditions at the origin $x=0$ that are $y(0) = 0$ and $y'(0) = 0$, and the condition at right boundary $y'(1) = 4e$.
The exact analytic solution is now $x^3 e^x$.

We have used our two PINNs variants, where the trial function chosen for hard-PINNs is now (following Lagaris prescription formula (1998)),
 \begin{equation}
      y_\theta (x) =  x^4 e + x^2(x-1)^2 \mathcal{NN} (x) .
    \end{equation}
The results obtained using vanilla-PINNs and hard-PINNs are plotted in Fig. 6 and Fig. 7 respectively. 
We use $N_c = 11$ collocation points.
The neural network parameters used are $2$ hidden layers with $20$ neurons per layer, with weight parameters $ \omega_{data}=1$, $\omega_d = 1 \times 10^{-2}$ (for vanilla variant).
$100$ points are used for the $MSE$ and errors evaluations.
 $\omega_{\mathcal{F}} = 5 \times 10^{-2}$, and a learning rate $\eta = 3  \times 10^{-4}$ are chosen. The training process is also stopped after $92000$ iteration steps (i.e. epochs).

     \begin{figure}[!t]
\centering
 \includegraphics[scale=0.34]{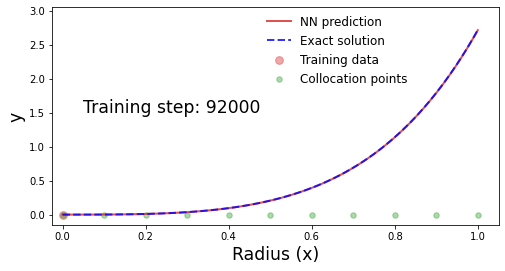}
  \includegraphics[scale=0.34]{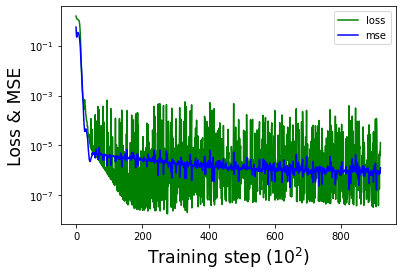}
   \includegraphics[scale=0.34]{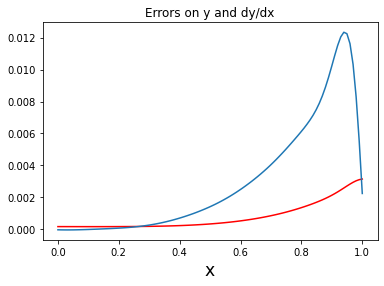}
  \caption{(Left panel) Predicted solution versus exact one using vanilla-PINNs obtained at the end of the training process for the third-order Emden-Fowler equation (second example).
  The radius $x$ values for collocation points at which the physical loss function is evaluated are indicated with the green circles on $x$ axis. (Middle panel) The history
  of the total loss function and $MSE$ function during the training (see text). (Right panel) Absolute errors on solution (red) and first order derivative (blue) at the end of the training.
   }
\label{fig6}
\end{figure}

     \begin{figure}[!t]
\centering
 \includegraphics[scale=0.34]{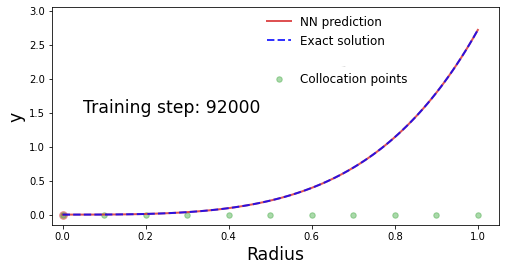}
  \includegraphics[scale=0.34]{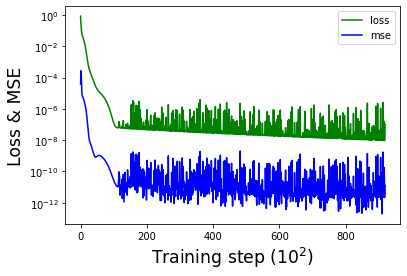}
   \includegraphics[scale=0.34]{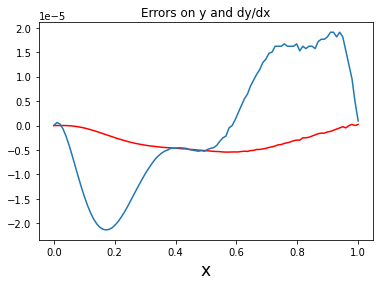}
  \caption{(Left panel) Predicted solution versus exact one using hard-PINNs obtained at the end of the training process for the third-order Emden-Fowler equation (second example).
  The radius $x$ values for collocation points at which the physical loss function is evaluated are indicated with the green circles on $x$ axis. (Middle panel) The history
  of the total loss function and $MSE$ function during the training (see text). (Right panel) Absolute errors on solution (red) and first order derivative (blue) at the end of the training.
   }
\label{fig7}
\end{figure}   

Contrary to the two previous examples, we can see that the precision reached with hard-PINNs is much better compared to the one obtained with vanilla-PINNs in this case.

\section{Application to fourth-order Lane-Emden-Fowler equations}

Many details about the fourth-order Lane-Emden-Fowler representative of multi-singular boundary value problems can be found
in Ali et al. (2022) and references therein. In the same above reference, one can also have a look on the comparison between different
numerical methods used to find solutions of such non trivial problems. In this work, we take only one example of this family of equations in order to 
illustrate how our PINNs technique already used above can be easily generalized for higher order equation. Other examples found in the literature
can be of course solved in the same way.

We consider the following equation,
    \begin{equation}
        \frac {d^4y} {dx^4 } +   \frac {12} {x } y''' +  \frac {36} {x^2} y'' +   \frac {24} {x^3} y' + 60 (7 - 18x^4 + 3 x^8) y^9 = 0 ,
         \end{equation}
for $x$ in the range $[0 : 1]$, and subject to the conditions at the origin $x=0$ that are $y(0) = 1$, $y'(0) = 0$, and $y''(0) = 0$, and $y'''(0) = 0$.
The exact analytic solution is now $ y = \frac {1} {\sqrt{1 + x^4}}$. The residual equation used in our PINNs variants is therefore,
    \begin{equation}
       x^3 \frac {d^4y} {dx^4 } +  12 x^2 y''' +  36x y'' +  24y' + 60x^3 (7 - 18x^4 + 3 x^8) y^9 = 0 .
         \end{equation}
The results obtained using vanilla-PINNs and hard-PINNs are plotted in Fig. 8 and Fig. 9 respectively. The trial function chosen for hard-PINNs is now 
 \begin{equation}
      y_\theta (x) =  1 - x^4 \mathcal{NN} (x) .
    \end{equation}
We use $N_c = 11$ collocation points.
The neural network parameters used are $2$ hidden layers with $20$ neurons per layer, with weight parameters $ \omega_{data}=1$, $\omega_d = 1 \times 1$ (for vanilla variant).
$100$ points are used for the $MSE$ and errors evaluations.
 $\omega_{\mathcal{F}} = 5 \times 10^{-3}$, and a learning rate $\eta = 5  \times 10^{-4}$ are chosen. The training process is also stopped after $150000$ 
 and $120000$ iteration steps (i.e. epochs) for the two variants respectively.

     \begin{figure}[!t]
\centering
 \includegraphics[scale=0.34]{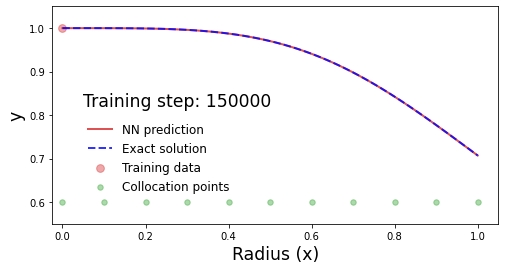}
  \includegraphics[scale=0.34]{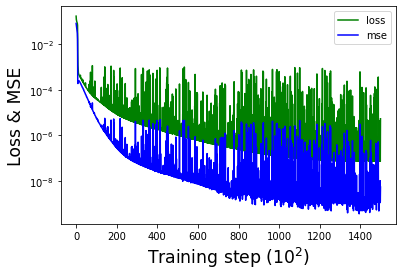}
   \includegraphics[scale=0.34]{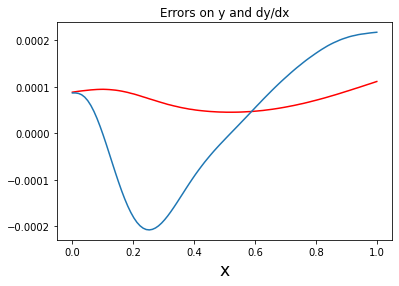}
  \caption{(Left panel) Predicted solution versus exact one using vanilla-PINNs obtained at the end of the training process for the fourth-order Lane-Emden-Fowler equation (se text).
  The radius $x$ values for collocation points at which the physical loss function is evaluated are indicated with the green circles on $x$ axis. (Middle panel) The history
  of the total loss function and $MSE$ function during the training (see text). (Right panel) Absolute errors on solution (red) and first order derivative (blue) at the end of the training.
   }
\label{fig8}
\end{figure}   

     \begin{figure}[!t]
\centering
 \includegraphics[scale=0.34]{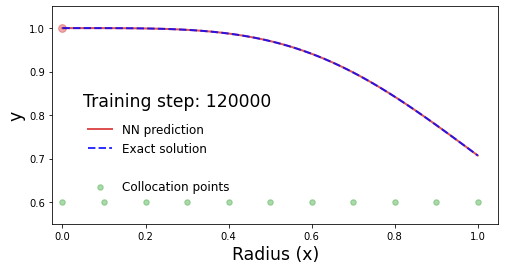}
  \includegraphics[scale=0.34]{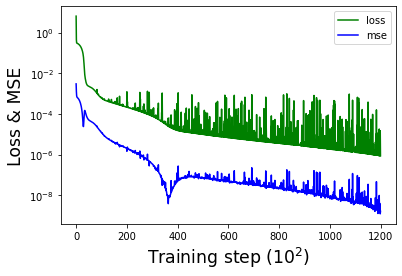}
   \includegraphics[scale=0.34]{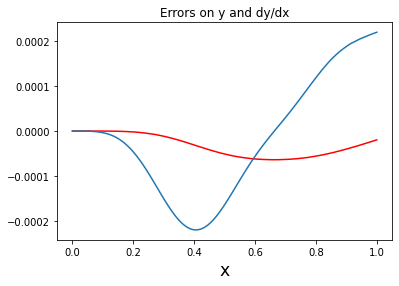}
  \caption{(Left panel) Predicted solution versus exact one using hard-PINNs obtained at the end of the training process for the fourth-order Lane-Emden-Fowler equation (se text).
  The radius $x$ values for collocation points at which the physical loss function is evaluated are indicated with the green circles on $x$ axis. (Middle panel) The history
  of the total loss function and $MSE$ function during the training (see text). (Right panel) Absolute errors on solution (red) and first order derivative (blue) at the end of the training.
   }
\label{fig9}
\end{figure}   

We see that the accuracy for the two variants is very similar for this case.

\section{Conclusions}

In this work, we show the potentiality of deep-learning techniques based on artificial neuronal networks and called Physics-Informed Neural
Networks (PINNs) to solve ODEs. These methods are conceptually simple, as they are based on the minimization of some defined functions.
Contrary to traditional numerical schemes, PINNs dot not require mesh discretization at which the differential operators are approximated.
For the two PINNs-variants presented in this paper, the so called loss functions include a physics-based loss function that represents the
equation residual evaluated at a relatively small dataset of collocation points.

With our vanilla-PINNs approach, the boundary conditions on the solution require to add corresponding constraints via
some training dataset and the associated training data loss function. When boundary conditions on solution derivatives are necessary,
the first and last collocation points can be used to impose the corresponding values. In this way, the boundary conditions may be
regarded as soft constraints imposed during the training process, as they are not exactly imposed.

With our hard-PINNs approach, the boundary conditions (on the solution and derivatives) can be imposed directly in the trial function
which approximates the solution. Contrary to the previous variant, the boundary conditions are hard constraints as they make the solution
and derivatives equal to the exact values by construction. As an advantage over vanilla-PINNs, this second variant requires a lower
number of hyper-parameters (i.e. the weights) and only one loss function, facilitating the convergence of the training. However, the obtention
of a well-behaved trial function is necessary.

When comparing the results obtained with the two variants in this work, we have found that the precision is in general similar but can be in some cases better with
the hard-PINNs. Whatever the variant used, PINNs methods are interesting advantages over traditional numerical integration methods.
 Once trained, the solution and derivatives can be quasi-instantaneously generated in the trained spatial domain. The solution
 obtained with our methods is valid over the entire domain without the need for interpolation (unlike RK tabular solutions). 
The eventual singularities are easily circumvented using our equation residual formulation. Moreover,  we do not need to transform the equation into a set of
 two/three/four first-order differential equations.
      
However, our results reveal some drawbacks. Indeed,
 the training process depends on a combination of many parameters like, the learning rate, the weights in the loss function, 
 and the architecture of the network, which determine the efficiency of the minimization. Consequently, a fine tuning is necessary to
find optimal parameter values which can therefore be computationally expensive.
 Even if the accuracy obtained in this work is excellent, PINNs seem to be potentially less accurate than classical methods
 where for example refining a grid (e.g. Runge-Kutta schemes) allows a precision close to the machine one. 
 This limitation is partly inherent to minimization techniques (see Press et al. 2007 and discussion in Baty 2023).
  
Anyway, we believe that PINNs are promising tools that are called upon to develop in future years, and ameliorations using self-adaptive techniques are expected
to improve the previously cited drawbacks (Karniadakis et al. 2021; Cuomo et al. 2022).      

\begin{acknowledgements}
The author thanks Emmanuel Franck, Victor Michel-Dansac, and Vincent Vigon (IRMA, Strasbourg), for associating him to the supervision of the Master2 internship
of Vincent Italiano in February-July 2022, which also gave him want to learn the PINNs technique.
 \end{acknowledgements}

 \section*{Data Availability}
Some of the Pytorch-Python codes used in this work will be made available on the GitHub repository at https://github.com/hubertbaty/PINNS-LEbis.

%

\begin{thebibliography}{}

 \bibitem[(Ali et al.(2022)]{Ali22} Ali,  K.K., Mehanna, M.S., Abdelrahman, M.A., Shaalan, M.A, 2022.
  Analytical and Numerical solutions for fourth order Lane-Emden-Fowler equation. Partial Differential Equations in Applied Mathematics 6, 100430

  \bibitem[Baty and Baty(2023)]{baty} Baty, H., Baty, L., 2023. Solving differential equations using physics informed deep learning: a hand-on tutorial with benchmark tests.
   Preprint https://arxiv.org/abs/2302.12260
   
   \bibitem[Baty(2023)]{baty} Baty, H., 2023. Modelling Lane-Emden Type Equations Using Physics-Informed Neural Networks.
   Accepted for publication in Astronomy and Computing. Preprint https://doi.org/10.1016/j.ascom.2023.100734
   
 \bibitem[Cuomo(2022)]{cuo21} Cuomo, S., Di Cola, V.S., Giampaolo, F., Rozza, G., Raissi, M., Piccialli, F., 2022.
 Scientific Machine Learning through Physics-Informed Neural Networks: Where we are and What's next.
 Journal of Scientific Computing 92, 88
       
  \bibitem[(Karniadakis et al.(2021)]{kar19} Karniadakis,  G.E., Kevrekidis, I.G., Lu, L, Perdikaris, P., Wang, S., Yang, L., 2021.
  Physics-informed machine learning. Nature reviews 3, 422-440
  
    \bibitem[(Khalid et al.(2014)]{kha14} Khalid, M., Sultana, M., Zaidi, F., 2014.
  Numerical Solution of Sixth-Order Differential equations Arising in Astrophysics by Neural Network.
  International Journal of Computer Applications 107,  0975887
    
   \bibitem[(Lagaris et al.(1998)]{lag98} Lagaris, E., Likas, A., DI Fotiadis, L., 1998. Artificial neural networks for solving ordinary and partial differential equations.
IEEE Transactions on Neural Networks 9(5), 987-1000
             
  \bibitem[Press(2007)]{pre07} Press, W.H., Teukolsky, S.A., Vetterling, W.T., Flannery, B.P., 2007.
 Numerical Recipes 3rd Edition
        
 \bibitem[Raissi et al.(2017)]{rai17} Raissi, M., Perdikaris, P., Karniadakis,  G.E., 2017. Physics Informed Deep Learning (Part I): Data-driven Solutions of Nonlinear Partial Differential Equations.
 Preprint https://doi.org/10.48550/arXiv.1711.10561
 
 \bibitem[Raissi et al.(2019)]{rai19} Raissi, M., Perdikaris, P., Karniadakis,  G.E.,  2019. Physics-informed neural networks: A deep learning framework
 for solving forward and inverse problems involving nonlinear partial differential equations.
 Journal of Computational Physics 378, 686-707

\bibitem[Verma and Kumar(2020)]{ver20} Verma, A., Kumar,  M.,  2020. Numerical solution of third-order Emden-Fowler type equations using artificial neural network technique.
 The European Physical Journal Plus 135, 751

    
\end{thebibliography}
%

\end{document}